\documentclass[10pt]{article}
\usepackage{amssymb,euscript,bbold}
\usepackage{amsfonts}
\usepackage{amssymb}
\usepackage{mathrsfs}
\usepackage{graphicx}
\usepackage{slashed}
\usepackage{amsthm}
\usepackage{amsmath}
\usepackage[latin1]{inputenc}
\newcommand{\be}{\begin{equation}}
\newcommand{\ee}{\end{equation}}
\newcommand{\ba}{\begin{eqnarray}}
\newcommand{\ea}{\end{eqnarray}}

\def\P{\mathbb{P}}

\def\R{\mathbb{R}}
\def\C{\mathbb{C}}

\begin{document}
\input{epsf}

\begin{flushright}
KCL-2019-56
\end{flushright}
\begin{flushright}
%{\sf \today}
\end{flushright}
\begin{center}
\Large{ Supersymmetry, Ricci Flat Manifolds and the String Landscape.}\\
\bigskip
\large{B.S. Acharya}\\
%\ft{}\\
\smallskip\normalsize{\it
Abdus Salam International Centre for Theoretical Physics, Strada Costiera 11, 34151, Trieste, Italy}\\

and

{\it Department of Physics, Kings College London, London, WC2R 2LS, UK}\\
\end{center}

%\renewcommand{\abstractname}{\sc Abstract}
%\begin{Abstract}
\bigskip
\begin{center}
{\bf {\sc Abstract}}
\end{center}
%\bigskip
%\normalsize
A longstanding question in superstring/$M$ theory is does it predict supersymmetry below the string scale? We formulate and discuss a necessary condition for this to be true; this is the mathematical conjecture that all stable, compact Ricci flat manifolds have special holonomy in dimensions below eleven. Almost equivalent is the proposal that
the landscape of all geometric, stable, string/$M$ theory compactifications to Minkowski spacetime (at leading order) are supersymmetric. For simply connected manifolds, we collect together a number of physically relevant mathematical results, emphasising some key outstanding problems and perhaps less well known results. 
For non-simply connected, non-supersymmetric Ricci flat manifolds we demonstrate that many cases suffer from generalised Witten bubble of nothing instabilities.

%\end{center}
%\end{abstract}
\newpage

\section{Introduction.}

Superstring/$M$ theory seems to be a promising framework for studying quantum gravity and the fundamental interactions of nature. A longstanding question has been, {\it does superstring/$M$ theory predict supersymmetry at low energies i.e. below the Planck, string , Kaluza-Klein or GUT scales?}. We will consider whether or not all stable compactifications to Minkowski spacetime are supersymmetric. In other words, is the low energy effective physics of superstring/$M$ theory supersymmetric just below the compactification scale?  In this paper we will address some necessary mathematical conditions for this to be the case, irrespective of the scale of supersymmetry breaking or the fine-tuning/hierarchy problem(s). We will see that the question is related to some challenging problems in differential geometry and global analysis such as the existence of compact, Ricci flat, Riemannian manifolds with generic holonomy groups.

We will be mainly interested in vacuum solutions of superstring/$M$ theory which are well approximated by the low energy supergravity approximation. In this approximation solutions consist of specifying a ten or eleven manifold, $M$, together with some specified geometric data. This data consists of specifying a Lorentzian metric on $M$ plus possibly other fields which will play little role in the sequel. The specified metric satisfies the Einstein equations (sourced by the other fields if present). This constitutes a classical approximation to a point in the String Landscape \cite{Swampland}. In principle, quantum effects will generate corrections to this classical solution but we will mostly concern ourselves with points in the String Landscape for which these corrections are small. We will thus be interested in points in the String Landscape described by classical geometry.

Motivated further by phenomenological considerations, we will only be interested in solutions which have some number of compact dimensions of space. 
For this one considers vacua in the string Landscape in which $M$ takes the form $X \times N$ where $X$ is {\it compact} and $N$ is a maximally symmetric Lorentzian spacetime. The metric on $X \times N$ is typically conformal to a product, $ds^2 = g(X) + g(N)$, where $g(N)$ is either the Minkowski, anti-de Sitter (AdS) or de Sitter metric. In the AdS case, $g(X)$ is typically a positive scalar curvature Einstein manifold, whereas in the Minkowski case it is typically a Ricci flat metric or conformal to one\footnote{The case of Hull-Strominger \cite{Hull-Strominger} metrics should also be considered, but these are Ricci flat at lowest order in the string tension. Note also that Type IIB flux compactifications are conformally Ricci flat \cite{Becker, GKP}.}. In the AdS case, Ooguri and Vafa have conjectured and provided evidence that the only stable AdS compactifications of superstring/$M$ theory are supersymmetric \cite{AdSSwamp}. The present paper discusses some analogous questions for Minkowski spacetimes. For a statistical approach to the problem of supersymmetry breaking see \cite{Douglas,Denef}.

When $g(N)$ is the Minkowski metric, $g(X)$ typically has zero Ricci tensor. Hence, we will mainly be concerned with compact, Ricci flat manifolds.  
If suitably stable under quantum corrections of all kinds, both perturbative and non-perturbative, then $(X, g(X))$ is a good approximation to a vacuum state of superstring/$M$ theory - a point in the String Landscape. This point in the landscape, at low energies, is described by an effective quantum field theory which includes gravity. The full theory is expected to be a consistent quantum theory of gravity coupled to matter. There is thus a correspondence between physical theories and compact, stable, Ricci flat manifolds\footnote{Of course, there are also stringy world-sheet conformal field theories and other less geometric physical theories which can also give rise to quantum theories in Minkowski spacetime, but here we focus on the geometric regime.}. One is thus interested in navigating and charting the landscape of all compact Ricci flat manifolds up to dimension  ten.

A basic structural result states that any compact Ricci flat manifold $(X, g(X))$ is isometric to a finite quotient of a product of compact Ricci flat manifolds. Thus, up to finite, discrete quotients, compact Ricci flat manifolds factorise. The {\it prime} factors in this decomposition are either compact and simply connected or they are flat tori. This factorisation is known as the Cheeger-Gromoll splitting theorem\footnote{Subsection 1.2 gives a sketch of the splitting theorem aimed at physicists.} \cite{Cheeger-Gromoll}.
\newpage
%\bigskip 
%\bigskip
\noindent
Therefore, we can consider the following four classes of {\it compact} Ricci flat manifolds:\\
Type A): simply-connected,\\ 
Type B): finite quotients of simply connected,\\ 
Type C): finite quotients of flat tori and \\
Type D): finite quotients of products of Type A) with tori.

%In some senses, Type C), the tori and their quotients are less interesting than the simply-connected, irreducible factors which appear in the factorisation, so we will mainly focus on irreducible Ricci flat manifolds to begin with.
%Later in the paper we will argue that many of the Type C) cases are physically unstable. We will consider cases of Type B) at the end of the paper and these are the most difficult cases to address and leave open several interesting questions.

%At the time of writing, we note that all known examples of compact, simply connected Ricci flat manifolds (in dimensions below eleven) are supersymmetric. The effective field theory %corresponding to these manifolds is a supersymmetric theory in Minkowski spacetime. This is due to the fact that all known examples have special holonomy. We will investigate the conjecture %that all stable compactifications to Minkowski spacetime are supersymmetric. For Ricci flat manifolds, this is the conjecture that all stable, compact, simply connected Ricci flat manifolds %have special holonomy.

Physically, Ricci flat manifolds have had a significant role. The seminal publication of \cite{CHSW} established what has turned out to be a long lasting relationship between superstring theory and the study of {\it Ricci flat manifolds with special holonomy groups}.
This relationship has been extremely fruitful, leading to remarkable results such as mirror symmetry for Calabi-Yau manifolds and enumerative properties such as Gromov-Witten, Donaldson-Thomas and Gopakumar-Vafa invariants, the construction of compact manifolds with holonomy groups $G_2$ and $Spin(7)$ and much more. Somehow, underlying all of this is an intuition that Ricci flat special holonomy manifolds are rather magical, rich mathematical objects. 

The present paper discusses the possibility (and a number of closely related conjectures) that the {\it the only} Ricci flat, compact, simply connected manifolds are special holonomy manifolds. 

This conjecture is interesting both physically and mathematically. Physically, if it were true, it would be a strong mathematical argument in favour of supersymmetry, irrespective of the hierarchy or naturalness problems. This is because Ricci flat, special holonomy manifolds admit parallel spinors and hence the corresponding vacuum solution is supersymmetric.
In particular, if the conjecture were true, it implies that {\it all consistent theories in the geometric regime with zero vacuum energy (at least at tree level) have supersymmetry below the Kaluza-Klein scale.} This would be a necessary condition for asserting that superstring/$M$ theory predicts supersymmetry below the Kaluza-Klein scale. Supposing that the conjecture were true, one might then be tempted to speculate that such a theorem extends beyond the geometric regime to all models of quantum gravity in Minkowski spacetime which arise from superstring/$M$ theory, implying that all non-supersymmetric Minkowski spacetime vacua are in the Swampland; this is in fact a conjecture that Banks has made from various different points of view \cite{Banks}.
Mathematically, regardless of whether it is true or not, we believe that that the question raises some interesting questions in differential geometry and analysis, such as the existence of Ricci flat metrics with generic holonomy in the compact setting.

In subsection 1.1 review the relationship between Ricci flat manifolds and supersymmetry and state our main conjecture. Subsection 1.2 gives a rough sketch of the splitting theorem for Ricci flat manifolds. In section 2 we consider stability questions which leads to a refinement of the conjecture. In section 3 we discuss the conjecture in the simply connected (Type A) category reviewing what is known in various dimensions, particularly four and eight. We discuss various obstructions to Ricci flatness and demonstrate ways in which Ricci flatness can fail. We show that most compact simply connected four manifolds do not admit a Ricci flat metric. In section 4 we demonstrate that many non-supersymmetric Ricci flat manifolds of Type C suffer from generalised Witten (bubble of nothing) instabilities. We argue that non-supersymmetric Type B Ricci flat manifolds may suffer a similar fate.
 
%{\bf This is a preliminary DRAFT from 16th May 2019!!} 

\subsection{\it Supersymmetry, Holonomy and Ricci flat backgrounds}

At a fundamental level, the relationship between physics and Ricci flat special holonomy manifolds comes from {\it supersymmetry}. Supersymmetric theories are mathematical models of physical systems whose underlying symmetry group is a supergroup. In most applications, this supergroup contains and extends the Poincare group. At the Lie algebra level, the symmetry algebra is a graded extension of the Poincare algebra. The grading distinguishes bosons (which are described by e.g. scalar, vector or tensor fields) from fermions (which are described by spinor fields). From a physical point of view, the key underlying property that a Ricci flat, special holonomy metric has, which distinguishes it from a generic metric, is that there exists a spinor field which is parallel with respect to the Levi-Cevita connection. In fact, the existence of such a parallel spinor is an equivalent definition of a Ricci flat special holonomy manifold. This is because parallel fields must be holonomy singlets and this requires holonomy reduction. A simply-connected compact manifold   admits parallel spinors with respect to the Levi-Cevita connection of a Riemannian metric if and only if that metric has special holonomy \cite{Wang}.

The Berger holonomy classification \cite{Berger}, which was later refined by Simons \cite{Simons} asserts that Riemannian metrics on a simply connected oriented manifold, $X_n$, of dimension $n$ must have one of the holonomy groups listed in Table 1. Hence, for our purposes, $g_X$ must belong to one of the Ricci flat cases, namely $SU(n)$ (aka Calabi-Yau), $Sp(k)$ (aka hyperK{\"a}hler) or one of the two exceptional cases, $G_2$ or $Spin(7)$. 

There remains however the open possibility, {\it labeled by the three question marks in Table 1}, that a compact, simply connected manifold with generic holonomy could be Ricci flat. In this case, such a manifold would provide, at least at leading order, a (possibly unstable, see below) {\it non-supersymmetric} background for superstring/$M$ theory. This is the key question considered in this paper.

A physical way of asking the same question is, {\it do compact, simply connected, non-supersymmetric Ricci flat manifolds exist?} What would the existence of such a generic holonomy, Ricci flat manifold mean? 

\begin{table}[h!]
  \begin{center}

    \label{tab:table1}
    \begin{tabular}{l|c|c|c} % <-- Alignments: 1st column left, 2nd middle and 3rd right, with vertical lines in between
      \textbf{Dimension} & \textbf{Holonomy Group} & \textbf{Ricci Flat} & \textbf{Parallel Spinor}\\
      $n$ & $Hol(g_X)$ & $R_{ij} = 0$? & $\nabla_g \eta =0$?\\
      \hline
      $n$ & $SO(n)$ & {\bf ???} & No\\
      $n=2k$ & $U(k)$ & No & No\\
      $n=2k$ & $SU(k)$ & Yes & Yes\\
      $n=4k$ & $Sp(k).Sp(1)$ & No & No\\
      $n=4k$ & $Sp(k)$ & Yes & Yes\\
      $n=7$ & $G_2$ & Yes & Yes\\
      $n=8$ & $Spin(7)$ & Yes & Yes\\
    \end{tabular}

    \caption{Riemannian Holonomy Groups, Ricci Flatness and Parallel Spinors.}

  \end{center}
\end{table}

We will thus consider the following basic conjecture:
\bigskip

\large
{\it Conjecture 1: Any compact, simply-connected, Ricci flat manifold has special holonomy.}

\normalsize

\subsection{The Structure of Ricci-flat manifolds.}

We conclude this introduction with a simple intuitive sketch, mainly aimed at physicists, for why the Cheeger-Gromoll splitting theorem, which provides our basic understanding of the structure of Ricci-flat manifolds, is true. This sketch essentially follows the paper of \cite{Fischer-Wolf} who provided an alternative proof in the Ricci flat case and their work in turn relies heavily on Bochner\textsc{\char13}s techniques \cite{Bochner}.

The basic idea is that on a compact Ricci flat manifold any harmonic 1-form is parallel with respect to the Levi-Cevita connection, $\nabla$. Similarly any Killing vector field is also parallel. This then provides a one-to-one correspondence between harmonic one-forms and (Abelian) isometries. Each such one-form corresponds to splitting off precisely one flat direction. 

If $\alpha$ is a harmonic one-form, with components $\alpha_i$, we have,
\be
\Delta\alpha_i = -g^{jk}\nabla_k \nabla_j \alpha_i + R_i^j \alpha_j = -g^{jk}\nabla_j \nabla_k \alpha_i = 0
\ee
Now, since $X$ is compact,
\be
\int_X g^{jk}\nabla_k \nabla_j \alpha_i\alpha^i = 0 = \int_X g^{jk}\nabla_k(\alpha_i\nabla_j\alpha^i + \alpha^i\nabla_j\alpha_i) 
\ee
From this it follows that
\be
\int_X \alpha^i g^{jk}\nabla_k \nabla_j \alpha_i = -\int_X \nabla_j\alpha_i\nabla^j\alpha^i
\ee
hence,
\be
\int_X \alpha^i\Delta\alpha_i = \int_X \nabla_j\alpha_i\nabla^j\alpha^i + R_{ij}\alpha^i \alpha^j
\ee
establishing that when the Ricci tensor vanishes, $\alpha$ is harmonic if and only if it is parallel. A parallel one-form is dual to a parallel vector field, which thus generates a shift symmetry in the metric and corresponds locally to isometrically splitting off an $S^1$ factor in the metric.
Similar calculations can be used to show that if $V$ is a Killing vector field and the Ricci tensor vanishes that $V$ is parallel and the corresponding one-form is harmonic.  Hence the correspondence is in both directions.

\section{The Stability of Ricci-flat manifolds.}

The background product metric $g(M)$ is Ricci-flat. However, physically this is not enough to ensure that one has a well-behaved background of superstring/$M$ theory. An important requirement is that the background is {\it stable}. This means both that the metric is stable under perturbations and also that the space should not decay into another space as time goes on. 

\subsection{\it Perturbative Stability} means that the metric $g(M)$ is stable under perturbations. At first order (i.e. tree level), this is the condition that, in a suitable gauge, the linearisation of $Ric(g(M))=0$ has a positive spectrum. In transverse, traceless gauge, this is equivalent to the spectrum of the so-called Lichnerowicz operator being semi-positive definite. Physically, this is the same as the absence of any negative mass squared modes whose presence would show that the metric, $g(M)$, is an unstable maximum of the potential: $V = -\int_M R dvol$. The Hessian of $V$ should be semi-positive. When $g(X)$ is a Ricci-flat metric of special holonomy, it is always perturbatively stable. In fact, this is guaranteed by supersymmetry: the existence of the parallel spinor(s), allows one to re-write the Lichnerowicz equation as the standard Laplacian acting on $p$-forms. Thus the condition of tree level perturbative stability is non-trivial only for the case of possible Ricci flat metrics with generic holonomy. 

Of course, there can in general be loop corrections to the potential, i.e. perturbative quantum corrections. We will not discuss these much here, but mention them for completeness. Since $V$ is the tree level potential it will generate its own loop corrections which could also destabilise the putative Ricci flat metric. If $g(M)$ has generic holonomy, then the low energy theory is non-supersymmetric and will not enjoy Bose-Fermi mass degeneracy. In this case one expects large quantum corrections to the (classically zero) cosmological constant. This is a straightforward argument which concludes that the vacuum state is not Minkowski in the non-supersymmetric case, but rather de Sitter or anti-de Sitter or perhaps some other cosmological solution. However, the absence of a solution to the cosmological constant problem implies that there are probably many subtleties lurking behind this argument so we will try not to rely on it.
See, for instance, \cite{NonSusyStrings} and references therein for some studies of non-supersymmetric compactifications of superstring theories, including computations of the effective potential.

\subsection{\it Non-perturbative Stability}
From a non-perturbative point of view, the statement is roughly that $g(M)$ is the metric with lowest energy amongst all metrics on manifolds {\it asymptotic to} $X \times  \R^{D-n,1}$. This is exactly as in the positive mass theorems\cite{SchoenYau,WittenStability}. Again, for special holonomy metrics, there is a (partial) stability result guaranteeing stability (under suitable conditions) \cite{Dai}. However, Ricci flat manifolds can be {\it unstable}.  An example was provided by Witten in \cite{BubbleofNothing},  though that particular example was non-simply connected: $M = S^1 \times {\R^{3,1}}$ with {\it odd} spin structure. We will return to the topic of non-simply connected Ricci flat manifolds later arguing that many of them have similar instabilities.

Physically, we are thus motivated to refine Conjecture 1:

\bigskip

\large
{\it Conjecture 2: Any stable Ricci flat metric on a compact, simply connected manifold has special holonomy.}

\normalsize

A perhaps more direct, physical, related conjecture, which includes the possibility of Type B) and C) manifolds is

\large
{\it Conjecture 3: Any absolutely stable {\it geometric} compactification of superstring/$M$ theory to Minkowski space is based on a special holonomy Ricci flat metric on a compact manifold.}

\normalsize

\smallskip

%Remark: note that we have restricted to simply connected manifolds (Type A) since we deal with the non-simply connected categories in section 4.

\bigskip

Notice that a counterexample to Conjecture 3 would be an absolutely stable non-supersymmetric string theory in asymptotically Minkowski spacetime and would rule out the possibility that superstring/$M$ theory predicts supersymmetry below the compactification scale. Conjecture 1 implies Conjecture 2 and Conjecture 2 is a necessary condition for proving Conjecture 3. Conjecture 3 is difficult since one requires control of quantum corrections to the potential in putative non-supersymmetric counterexamples such as \cite{NonSusyStrings}. 
A proof of Conjecture 1 or Conjecture 2 is necessary to demonstrate  that geometric compactifications of superstring/$M$ predict supersymmetry below the compactification scale.

\section{Obstructions to Ricci Flat Metrics}

As we will see, there can exist obstructions to the existence of Ricci flat metrics. The simplest of these are topological obstructions. However, by considering explicit attempts to construct Ricci flat metrics one can encounter more subtle obstacles, as we will review. These are certainly worthy of more detailed study. In this section we mainly focus on the simply connected category (Type A).

\subsection{\it Topological Obstructions}

In dimensions one, two and three, Ricci flatness is equivalent to local flatness, in which case there are no compact, simply connected examples. Dimension four is the lowest dimension in which it is possible to have a compact, simply connected Ricci flat manifold.  
In dimension four, Conjecture 2 would imply that $K3$ is the only simply-connected compact four manifold which admits stable Ricci flat metrics. Dimension four is also a setting in which many useful results are known, such as Freedman\textsc{\char13}s theorem \cite{Freedman} which classifies compact, simply-connected, topological 4-manifolds and a wealth of results which followed from Donaldson\textsc{\char13}s gauge theoretic approach \cite{Donaldson} and, then, later, the Seiberg-Witten equations \cite{SW}.

The topological types of simply-connected, compact four-manifolds are classified as follows:
If $X$ is spin (and satisfies the 11/8 conjecture), then it is a connected sum of $m$ $K3$ s and $n$ copies of $S^2 \times S^2$. In the non-spin case, it is a connected sum of $p$ ${\bf CP^2}$s and $q$  $ \bar{\bf CP^2}$s. Let us denote these manifolds as $X^4(m,n)$ and $Y^4(p,q)$. In dimension four, our conjecture is the statement that out of all of these four-manifolds, only the case $X^4(m=1,n=0)=K3$ admits stable, Ricci flat metrics. Donaldson\textsc{\char13}s work also led to the demonstration of the existence of many exotic smooth structures on these manifolds. In the case of $K3$, Ricci flat metrics exist only in the standard smooth structure that it inherits as a complex manifold. In any case, exotic smooth structures will not play much of a significant role in what follows, though it is important to bear in mind that a given four manifold could admit several, even many exotic smooth structures.

For lots of values of $(m,n)$ and $(p,q)$ one can prove that Ricci flat metrics do not exist. In fact, there can be topological obstructions. Perhaps the most well known of these is the Hitchin-Thorpe inequality \cite{HitchinThorpe} which is an inequality between the Euler number $\chi(X)$ and its signature $\tau(X)$ :

\bigskip

Theorem (Hitchin-Thorpe): {\it An oriented, Einstein, 4-manifold, $(X,g)$ satisfies}
$2\chi(X) \geq 3 |\tau(X)|$. 

\bigskip

Hence, every Ricci flat 4-manifold must satisfy this topological condition. Combining this with the above, it is clear that if $q$ is sufficiently large for fixed $p$, $Y^4(p,q)$ violates the Hitchin-Thorpe inequality and, hence, no Ricci flat metric can exist.
One can extend these ideas by making use of Seiberg-Witten theory. LeBrun in particular has shown that there exist manifolds which satisfy the Hitchin-Thorpe inequality, but nonetheless do not admit Einstein metrics \cite{LeBrun}.

When $X$ is a spin manifold, a similar argument can be obtained using spinors and the Lichnerowicz-Weitzenbock formula and was first exploited by Lichnerowicz and then by Hitchin \cite{Lichnerowicz, Hitchin}. Let $D_g$ denote the Dirac operator associated with the metric $g_X$. $D_g$ acts on sections of the spin bundle of $X$. The square of $D^g$ on $(X,g)$ 
satisfies:
\be
D_g^2 = \nabla_g^2 + {1 \over 4} R_g
\ee
where $\nabla_g$ is the Levi-Civita connection and $R_g$ the scalar curvature. 

Clearly then, if $g$ is a Ricci flat metric, $R_g=0$, so zero modes of $D_g$ are in fact parallel spinors.
Since any Ricci flat metric which admits parallel spinors actually has special holonomy, a Ricci flat manifold of generic holonomy necessarily has no harmonic spinors and, hence, zero Dirac index. Therefore, we see that

\bigskip

Theorem (Hitchin): {\it a compact manifold with non-zero Dirac index can never admit a Ricci flat metric of general holonomy.}

\bigskip

One can exploit this argument to construct large numbers of manifolds which can not admit Ricci flat metrics, simply by taking connected sums to produce manifolds with non-zero Dirac index. 
For example, in dimension four, for
$X^4(m,n) = mK3\#n(S^2\times S^2)$, we have that the Dirac index, $\hat{A}(X^4(m,n)) = 2m$. Hence, we obtain:

\smallskip

Corollary: {\it Let $X^4(m,n)$ be the 4-manifold obtained by taking the connect sum of $m$ K3 s and $n$ copies of $S^2 \times S^2$. For all $m\geq 1$ and $n\geq 1$ 
$X^4(m,n)$ cannot admit a Ricci flat metric. Similarly, $X^4(m \geq 2, 0)$ cannot admit a Ricci flat metric.}

\bigskip

Two interesting cases are worth noting upon here. Firstly, $X^4(1,n)$ admits Ricci flat metrics if and only if $n=0$. This is because, $\hat{A}(X^4(1,n)) = 2$ (independent of $n$) but having two parallel spinors implies that a simply connected 4-manifold must be $K3$. 
This means that one cannot glue $S^2 \times S^2$ into Ricci flat $K3$ manifolds whilst preserving Ricci flatness. This brings us to our second point, which is apparently a fairly longstanding unsettled\footnote{Indeed, unsettling.} question: 

\bigskip

\large
{\it Question: does $S^2 \times S^2$ admit a Ricci flat metric? }

\normalsize

\bigskip

We hope that this paper will serve to stimulate further interest in this question (and also the question for $X^4(0,n)$).
Perhaps using the fact that $X^4(1,1)$ does not admit a Ricci flat metric, by taking a suitable ansatz one can make progress on this question, since it is the glueing in of 
$S^2\times S^2$ into $K3$ that destroys Ricci flatness.
A related, interesting result is due to Kapovitch and Lott, who have shown that compact 4-manifolds with non-zero Dirac index, volume bounded below, diameter bounded above and suitably small Ricci curvature must in fact be $K3$ surfaces \cite{Kapovitch-Lott}.

%{\bf This is a preliminary DRAFT from 16th May 2019!!} 

One can clearly extend these sorts of arguments to higher dimensions. For instance, In dimension eight, the Ricci flat special holonomy groups are $Sp(1) \times Sp(1), Sp(2), SU(4)$ and $Spin(7)$. The index of the Dirac operator on these manifolds is, resp. $4,3,2,1$ and examples exist realising all four possibilities. Therefore, 

Theorem (Futaki, \cite{Futaki}): {\it Let $X^8$ be a simply connected, compact 8-manifold with non-zero Dirac index. Then $X^8$ can admit a Ricci flat metric if and only if the index is $1,2,3,4$ and such a metric has special holonomy.}

Corollary: {\it Let $X^8$ be a compact manifold which admits a metric with holonomy $Spin(7)$ and let $X^8_p = \#pX^8$ be the manifold obtained by taking the connect sum of $X^8$ with itself, $p$ times. When $p \geq 5$ $X^8_p$ cannot admit a Ricci flat metric.}

It would be very interesting to see how many homotopy types of 8-manifolds can be realised by taking connected sums using only special holonomy manifolds.

For completeness we extend these statements to $4k$ dimensions when $k\geq 3$. In these dimensions, neglecting Riemannian products for simplicity, the only possibilities for Ricci flat special holonomies are $Sp(k)$ and $SU(2k)$. Manifolds with these holonomy groups have Dirac index $k+1$ and $2$ respectively. Therefore, any (non-product) compact, simply connected $4k$-manifold whose Dirac index is non-zero cannot admit a Ricci flat metric unless the index is $k+1$ or $2$, in which case the only Ricci flat metrics will have special holonomy. 

Unfortunately, the Dirac index is non-zero only in 4$k$-dimensions, so these arguments are not of much use in the physically interesting cases of dimensions six and seven. However, in dimensions $8k+1$ and $8k+2$, there is a refinement of the Dirac index \cite{alpha} which can also obstruct the existence of Ricci flat metrics. This $\alpha$-invariant is the number of Dirac zero modes mod 2 in odd dimensions and the number of positive chirality zero modes mod 2 in even dimensions. Thus, when $\alpha$ is non-zero a Ricci flat metric must have special holonomy.
Baraglia \cite{Baraglia} has recently shown that many examples of compact, simply connected K\"{a}hler manifolds of real dimension ten have non-zero $\alpha$-invariant. When $\alpha$ is non-zero for a compact, simply connected, Ricci flat, ten manifold then that manifold must be a Calabi-Yau with holonomy equal to $SU(5)$. But most examples in \cite{Baraglia} have non-zero first Chern class and hence are not Calabi-Yau. For example, Baraglia shows that if $X_{10}$ is a degree 7 (mod 16) or 9 (mod 16) hypersurface in ${\C\P^6}$, then $\alpha=1$, but only the lowest degree case is Calabi-Yau. These manifolds therefore cannot admit Ricci flat metrics. We also note that in nine and ten dimensions, there exist exotic spheres, $\Sigma^{9,10}$ with $\alpha(\Sigma^{9,10})=1$ \cite{Hitchin}. So by taking a connect sum with an appropriate odd or even number of the $\Sigma^{9,10}$ one can make many topological manifolds with non-zero $\alpha$. 

\subsection{\it Explicit Constructions}

One could also attempt to prove the existence of Ricci flat metrics with generic holonomy on simply connected manifolds by explicit gluing constructions.
For instance, by gluing in a non-compact, known Ricci flat manifold into a known Ricci flat manifold with boundary. 
Following a suggestion of Page, Brendle and Kapouleas considered what might be termed a {\it non-supersymmetric} Kummer construction in \cite{BK}. Here, the idea is to modify the standard Kummer construction of Ricci flat $K3$ manifolds. 

In the classic Kummer construction, one begins with a singular, locally flat orbifold, ${T^4}/{\bf Z_2}$, which has 16 singular points, locally modelled on the origin in ${\R^4}/{\bf Z_2}$. To prove the existence of a hyperK{\"a}hler metric on a smooth $K3$ manifold, one removes a small neighbourhood of the singular set and glues in a very small Eguchi-Hanson (EH) manifold to make a simply connected 4-manifold diffeomorphic to $K3$. With a suitable gluing ansatz, one can assume that the metric on the $K3$ is very well approximated by the special holonomy Eguchi-Hanson metrics near these glued in regions. Finally, some perturbation theory and non-trivial analysis is used to demonstrate the existence of the supersymmetric, Ricci flat metric. This was originally proven in \cite{Topiwala,LeBrun-Singer}.
Donaldson also has an excellent, recent exposition of this construction \cite{DonaldsonGluing}. A crucial point about the geometry and topology of this example is that the Eguchi-Hanson spaces must all be glued in with the same orientation in order to produce a manifold whose signature is 16 (or minus 16). However, as real manifolds one can glue them in with either orientation to obtain a topological 4-manifold which will not be diffeomorphic to a $K3$ surface unless they all have the same orientation. This process breaks supersymmetry, because an Eguchi-Hanson with the original orientation has two parallel spinors which are sections of the positive chirality spin bundle $S^+$, but those of its orientation reversed reflection are sections of $S^-$. 
Brendle and Kapouleas consider the case with eight Eguchi-Hanson spaces and eight anti-Eguchi-Hanson spaces arranged in a symmetric checkerboard configuration amongst the 16 fixed points. They concluded that there is no Ricci flat metric at second order in perturbation theory close to the Kummer limit. Of course, this does not prove there is not a Ricci flat metric in some other region of the space of metrics, but it is physically striking that, even though the EHs and anti-EHs are localised in {\it arbitrarily} small regions and are {\it arbitrarily} far apart from one another, that they sense each others presence via the long range induced potential on the space of metrics.

It would certainly be very interesting to analyse the less symmetric versions of this construction further and also a related simpler, non-compact version: the locally flat orbifold $({\R^3 \times S^1})/{\bf Z_2}$ which has two fixed points both modelled locally on ${\R^4}/{\bf Z_2}$ could also be desingularised by gluing in Eguchi-Hanson$'$s of opposite orientation. Does this admit a Ricci flat metric? One can also consider higher dimensional generalisations of this idea. For example, compact Calabi-Yau manifolds often have points in the space of Ricci flat metrics at which isolated singularities develop, such as conifold singularities. If one can view the smooth nearby Ricci flat Calabi-Yau metrics as arising from a gluing construction, one could then produce potential non-supersymmetric examples by changing the orientations and/or complex structure of the non-compact model metric that one uses to glue in.

\section{\it Non-simply Connected Ricci flat Manifolds and Witten Instabilities.}

We now consider non-simply connected compact Ricci flat manifolds. There are essentially two cases to consider: (Type C) finite quotients of tori or (Type B) finite quotients of compact simply-connected Ricci flat manifolds. We consider Type C first.

\subsection{\it Compact Flat Manifolds.} All such manifolds are finite quotients of flat $T^n$ by a subgroup, $\Gamma$, of the Euclidean group $E(n) \cong O(n) \ltimes \R^n$. We will restrict our attention to the oriented case, so the rotations in $\Gamma$ are in $SO(n)$. Since $T^n/\Gamma$ is locally flat, non-trivial holonomies are generated by $\Gamma$ and, in fact, $Hol(T^n/\Gamma) = Rot(\Gamma)$, where $Rot$ is the natural projection $Rot: E(n) \rightarrow SO(n)$ which forgets translations. Since the holonomy group is non-trivial, we can ask: when is
$T^n/\Gamma$ supersymmetric? The answer is that it must admit a parallel spinor. This was addressed in dimension three in \cite{Pfaeffle}. In dimension three there are six compact orientable 3-manifolds and by explicitly calculating the spectrum of the Dirac operator in all of the spin structures on all of these manifolds, it was concluded that $T^3$ with the standard spin structure is the only manifold with a parallel spinor. In this sense $T^3$ with its standard spin structure is the only supersymmetric, compact oriented Ricci flat 3-manifold. There is a natural explanation of this fact: $Spin(3)$ has no non-trivial subgroup which preserves spinors. 

In higher dimensions however, $Spin(n)$ contains subgroups which preserve spinors and these subgroups must be contained precisely in one of the Ricci flat special holonomy groups in Table 1. Hence, the natural extension of this result is that $T^n/\Gamma$ admits parallel spinors precisely when
$Rot(\Gamma)$ belongs to one of the Ricci flat special holonomy groups. Since these cases are supersymmetric, we now focus on the non-supersymmetric cases. We will show that, in many cases there is a non-perturbative instability of the type first discussed by Witten in \cite{WittenStability}.

Witten considered the following non-compact Ricci flat metric on $\R^2 \times S^3$:
\be
ds^2 = \frac{dr^2}{(1-R^2/r^2)}+r^2(d\theta^2+cos^2\theta d\Omega_2^2) + (1-R^2/r^2)R^2d\phi^2
\ee
where $r \in |R,\infty)$, $R$ is a positive constant, $\theta \in (0,\pi)$, $d\Omega_2^2$ is the round metric on the unit two-sphere and $\phi \in (0,2\pi)$. The bracket multiplying $r^2$ in the second term is the round metric on the three sphere, written as a {\it (co-)sine cone}.

The metric is obtained as an analytic continuation of the five dimensional Lorentzian Schwarzschild metric, which is the explanation of the fact that $\phi$ is a coordinate on a unit circle. 
A further analytic continuation $\theta \rightarrow i t$ yields a Lorentzian metric describing Witten$'$s bubble of nothing. At $r \rightarrow \infty$ the geometry is flat $\R^{3,1} \times S^1$,
where the circle has radius $R$. A coordinate transformation to standard Minkowski coordinates reveals that in the complete manifold, what is ordinarily the inside of the lightcone does not actually exist since the geometry ends at $r=R$. Hence, the metric describes a bubble which is expanding faster and faster and asymptotically, at $t\rightarrow\infty$ approaches {\it nothing}, hence the terminology.

From a mathematical point of view, one can view the manifold as a spacetime, asymptotic to flat $\R^{3,1} \times S^1$ but also with zero energy \cite{WittenStability}. A crucial fact, however, is that, since $\R^2\times S^3$ has a unique spin structure, the geometry singles out a spin structure on the circle at infinity. In fact, this is the odd (or bounding) spin structure, because of the first analytic continuation. Equivalently, this is because the circle collapses in the geometry and thus bounds a disc. Therefore, $\R^{3,1} \times S^1$ is unstable in this sense, but only with the odd spin structure. This is in keeping with our conjecture that non-supersymmetric Ricci flat manifolds are unstable, because the odd spin structure is naively incompatible with supersymmetry\footnote{see \cite{GregNeil} for discussions about supersymmetry in these contexts.} and does not admit a parallel spinor. We now discuss a higher dimensional analogue of this phenomenon. In particular,
we will explain how compact flat 3-manifolds can decay via a similar mechanism.

\subsubsection{Compact, Orientable Flat 3-manifolds and Witten Instabilities}.

In dimensions one and two the only orientable, compact examples are $S^1$ and $T^2$. In dimension three, there are six classes of examples. These six classes can be uniquely specified by their holonomy groups. Let ${\R^3}$ have coordinates $(x_1 , x_2, x_3)$ and denote by $R_i(\theta)$ the $SO(3)$ rotations by angle $\theta$ about the $x_i$ axis.
Further, let $(R_i(\theta), \vec{b})$ denote the elements of the Euclidean group given by rotations about the $x_i$-axis combined with a translation by $\vec{b}$ so that $\vec{x} \rightarrow R.\vec{x} + \vec{b}$. Then the following table describes all six classes (up to moduli which have been fixed for convenience), $G1$ to $G6$. The second column specifies the lattice defining the covering $T^3$, the third gives the generators of $\Gamma$ which acts freely on the $T^3$ and the final column the holonomy group.

\begin{table}[h!]
  \begin{center}

    \label{tab:table1}
    \begin{tabular}{l|c|c|c} % <-- Alignments: 1st column left, 2nd middle and 3rd right, with vertical lines in between
      \textbf{Manifold} & \textbf{Lattice} & \textbf{$\Gamma$} & \textbf{Holonomy}\\
      \hline
      $G1$ & $a_1 = (1,0,0);a_2 = (0,1,0);a_3 = (0,0,1)$ & 1 & {\bf 1}\\
      \hline
      $G2$ & $(1,0,0);(1,1,0);(0,0,1)$ & $\alpha = (R_3(\pi),{a_3 \over 2})$ & ${\bf Z_2}$\\
      \hline
      $G3$ & $(1,0,0);(-{1\over 2},{\sqrt{3}\over 2},0);(0,0,1)$ &$\alpha = (R_3({2\pi \over 3}),{a_3 \over 3})$ & ${\bf Z_3}$\\
            \hline
$G4$ & $(1,0,0);(0,1,0);(0,0,1)$ &$\alpha = (R_3({\pi \over 2},{a_3 \over 4})$& ${\bf Z_4}$\\
      \hline
      $G5$ & $(1,0,0);({1\over 2},{\sqrt{3}\over 2},0);(0,0,1)$ & $\alpha = (R_3({2\pi \over 6}),{a_3 \over 6})$ & ${\bf Z_6}$\\
      \hline
      $G6$ &  & $\alpha = (R_3(\pi),{a_3 \over 2})$ & \\
$ $ & $(1,0,0);(0,1,0);(0,0,1)$ & $\beta = (R_1(\pi),{a1 + a_2 \over 2})$ & ${\bf Z_2 \times Z_2}$\\      
$ $ & & $\gamma = (R_2(\pi),{a1 + a_2 +a_3 \over 2})$ & \\      

      \hline
    \end{tabular}

    \caption{Compact, orientable flat 3-manifolds.}

  \end{center}
\end{table}

A crucially important aspect of Witten\textsc{\char13}s bubble of nothing instability was the spin structure of the circle. The spin structures of these six 3-manifolds will also play a crucial role here. These have been classified in \cite{Pfaeffle} and we reproduce those results here for completeness. A spin structure on these flat 3-manifolds is essentially a homomorphism from the lattice generators and $\Gamma$ to $Spin(3) = SU(2)$ whose projection down to $SO(3) = {SU(2) \over {\bf Z_2}}$ gives the appropriate rotations.
Each of the examples $G2$ through $G5$ have the property that $\alpha^N = a_3$, for $N=2,3,4,6$. In example $G6$ we have $\alpha^2=a_3$ and $\beta^2=a_1$ and $\gamma^2=a_2$.
These relations fix the choices of allowed spin structures to be those given in Table 3.

\begin{table}[h!]
  \begin{center}

    \label{tab:table1}
    \begin{tabular}{l|c|c|c} % <-- Alignments: 1st column left, 2nd middle and 3rd right, with vertical lines in between
      \textbf{Manifold} & \textbf{Lattice} & \textbf{$\Gamma$} & \textbf{Number}\\
      \hline
      $G1$ & $a_1 \rightarrow \delta_1;a_2 \rightarrow \delta_2;a_3 \rightarrow \delta_3$ & $\delta_i \in \pm {\bf 1}$ & 8\\
      \hline
      $G2$ & $a_1 \rightarrow \delta_2;a_2 \rightarrow \delta_3;a_3 \rightarrow -{\bf 1}; \alpha \rightarrow \delta_1 \hat{\alpha}$&$\delta_i \in \pm {\bf 1}$& 8 \\
      \hline
      $G3$ & $a_1 \rightarrow {\bf 1};a_2 \rightarrow {\bf 1};a_3 \rightarrow \delta_1; \alpha \rightarrow {- \delta_1} \hat{\alpha}$ &$\delta_i \in \pm {\bf 1}$&2\\
            \hline
$G4$ & $a_1 \rightarrow \delta_2;a_2 \rightarrow \delta_2;a_3 \rightarrow -{\bf 1}; \alpha \rightarrow \delta_1 \hat{\alpha}$&$\delta_i \in \pm {\bf 1}$&4\\
      \hline
      $G5$ & $a_1 \rightarrow {\bf 1};a_2 \rightarrow {\bf 1};a_3 \rightarrow -{\bf 1}; \alpha \rightarrow \delta_1 \hat{\alpha}$ &$\delta_i \in \pm {\bf 1}$&2 \\
      \hline
      $G6$ & $a_1 \rightarrow -{\bf 1};a_2 \rightarrow -{\bf 1};a_3 \rightarrow -{\bf 1}; \alpha \rightarrow \delta_1 i \sigma_3 ;\beta \rightarrow \delta_3 i\sigma_2; \gamma \rightarrow \delta_2 i \sigma_1$ & $\delta_1\delta_2\delta_3={\bf 1}$ & 4\\
      \hline
    \end{tabular}

    \caption{Spin structures on compact, orientable flat 3-manifolds.}

  \end{center}
\end{table}

In the $G2$ through $G5$ cases, $\hat{\alpha}$ denotes the diagonal $SU(2)$ matrix with entries $(e^{\pi i \over N},e^{-\pi i \over N})$ which satisfies $\hat{\alpha}^{2N}={\bf 1}$. In the $G6$ case, the three elements of the holonomy group map to $i$ times the three Pauli matrices. In total, there are 28 choices of spin structures on $G1-G6$. Our conjecture would assert that all of these are unstable except $T^3$ with the standard spin structure. Notice that in all 28 cases except two, the spin structures on the 3-manifold are inherited from an {\it odd} (i.e. non-standard) spin structure on $T^3$. This is due to the spin structure $a_3 \rightarrow {\bf -1}$ in all these cases. This is exactly the direction in which the generators of $\Gamma$ have a translation along $T^3$. The remaining two cases are the standard spin structure on $T^3$ which is supersymmetric and $G3 = T^3/{\bf Z_3}$ in the case $a_3 \rightarrow {\bf 1}$.

Consider now the following Ricci flat metric on $X^7 = \R^2\times S^3 \times T^2$:
\be
ds^2 = \frac{dr^2}{(1-R^2/r^2)}+ r^2 (d\theta^2 +cos^2\theta d\Omega_2^2)
+ (1-R^2/r^2)R^2dx_3^2 + dx_1^2 + dx_2^2
\ee
with $x_1$ and $x_2$ coordinates on $T^2$. In this metric, it is natural that the coordinates $(r,x_3)$ are polar coordinates on $\R^2$ minus a disc of radius $R$. 
This metric mediates the decay of $\R^{3,1} \times T^3$ but with odd spin structure in the $x_3$ direction. By permuting the $x_i$ accordingly, we get {\it a metric describing the decay of $T^3$ with any odd spin structure}.

Furthermore, the metric is also invariant under $\Gamma$ transformations. Therefore, with the appropriate identifications, the metric above also describes the decay of ${T^3/\Gamma}$ in all 26 out of 28 cases in which the spin structure of the covering $T^3$ is not the standard one.
We can also think of it as a metric on $(\R^2\times S^3 \times T^2)/{\Gamma}$, which can be regarded as a flat bundle over the singular orbifold $T^2/{\Gamma}$ with generic fibre 
$\R^2 \times S^3$ and fibres $(\R^2 \times S^3)/{\Gamma}$ over the fixed points of $\Gamma$ acting on $T^2$. In other words, these manifolds can be regarded as $T^2/\Gamma$ families of Witten\textsc{\char13}s original example. This seems to establish that 26 out of the 27 possible non-supersymmetric spin structures on compact flat 3-manifolds are unstable in this sense.

There is a small subtlety however when $\Gamma \neq {\bf 1}$. Since the circle in the bubble of nothing metrics collapses, and $\Gamma$ rotates this circle, $\Gamma$ has fixed points on $\R^2\times S^3 \times T^2$. For instance, for the case $\Gamma = {\bf Z_2}$ the four fixed points on $T^2/{\bf Z_2}$ lead to four fixed $S^3$'s. Near each of these four orbifold singularities, $X^7$ looks like $\R^4/{\bf Z_2} \times S^3$. Such codimension four singularities are  familiar in superstring and $M$ theory where they represent the presence of extra light degrees of freedom. In the case of $M$ theory for instance, there are $SU(2)$ gauge fields localised on the four $S^3$'s. Thus, these more general bubbles of nothing have some additional novel features that might have further consequences. 

All of the flat backgrounds, $T^3/{\Gamma}$, can be studied as exact worldsheet conformal field theories following \cite{Dixon} in which case the exact mass spectrum of string states can be computed. At small radii, of order the string length, they all contain tachyonic states and perhaps this is consistent with the instabilities uncovered here.

In higher dimensions it is clear that this phenomenon should generalise and we would therefore conclude that many non-supersymmetric compact flat manifolds are unstable. It would certainly be interesting to investigate this further and there are a wealth of examples since the number of compact flat manifolds grows rapidly with dimension. Clearly though, whenever $\Gamma$ contains generators, $\alpha$, of even order, the corresponding spin structures must be inherited from non-standard spin structures on the covering $T^n$ and, hence, we expect a generalised Witten instability.

We finally go on to discuss finite quotients of compact, simply connected Ricci flat manifolds, that is Type B.

\subsection{\it Quotients of Compact Simply connected Ricci flat manifolds}.

We now discuss Type B. All such manifolds are of the form $M \cong X/\Gamma^\prime$, where $X$ is compact and simply-connected and $\Gamma^\prime$ is a freely acting finite group. If $X$ is such that it admits a Ricci flat metric of generic holonomy, then it is treated in the previous section. The new phenomenon that we consider here is when $X$ has special holonomy and parallel spinors which are however {\it not} $\Gamma^\prime$-invariant, so that $M$ does not admit parallel spinors and
$X/\Gamma^\prime$ is not supersymmetric.  

The classic example of this phenomenon are the Enriques surfaces, $S$.
These are all diffeomorphic to $K3/{\bf Z_2}$ for some special $K3$ surfaces. Thus, if one has a $K3$-surface with a Ricci flat metric admitting a freely acting ${\bf Z_2}$ symmetry, the metric inherited by the Enriques surface $S$ will also be Ricci flat. Unlike the dimension three examples discussed above though, $S$ is not a spin manifold. However, presumably string theory on $S$ makes sense, when viewed as an orbifold conformal field theory. We can give one such description here, which is a flat orbifold limit of $S = {T^4}/({\bf Z_2 \times Z_2})$. We emphasise, that whilst superficially similar to the flat examples considered previously, this example is not of Type C, since the fundamental group of Type B examples is finite, unlike Type C manifolds.

Denoting coordinates on $T^4$ as before as $(x_a)$ $a=1,2,3,4$, and assuming a square lattice with the periodicities $x_a \sim x_a+2\pi$, the action of ${\bf Z_2 \times Z_2}$ is given by
\be
\begin{aligned}
\alpha: (x_1,x_2,x_3,x_4) & \rightarrow (-x_1-\pi,-x_2,x_3+\pi, x_4) \\
\beta: (x_1,x_2,x_3,x_4) & \rightarrow (-x_1,-x_2,-x_3,-x_4)
\end{aligned}
\ee

Notice that $\alpha$ and $\alpha\beta=\beta\alpha$ both act freely on $T^4$.

As an orbifold conformal field theory, there are tachyonic states in the small radius limit, so there is certainly an instability at small volume. At large volume we can try and follow the discussion of the previous subsection. It would be very interesting to study this and the examples of the previous section in more detail using the worldsheet conformal field theory.

One can view this particular, singular Enriques surface $S$ as a quotient by $\alpha$ of a Kummer K3 = ${T^4}/{\bf Z_2}$. But, one can also consider it as a quotient of 
${T^3}/{\bf Z_2} \times S^1$ by $\beta$ and, since ${T^3}/{\bf Z_2}$ has the Witten instability we discussed above, it seems likely that 
\be
ds^2 = \frac{dr^2}{(1-R^2/r^2)}+ r^2 (d\theta^2 +cos^2\theta d\Omega_2^2)
+(1-R^2/r^2)R^2dx_3^2 + dx_1^2 + dx_2^2 + dx_4^2
\ee
describes the decay of $S$. Formally, this is true, however, our model for $S$ is singular and one would like to understand whether or not a suitable gluing construction demonstrates that 
smooth Enriques surfaces obtained by a generalised Kummer construction are unstable in this sense. The singularities here are different in nature to those of the previous subsection since $\beta$ acts with fixed points on the circle which collapses in the bubble of nothing. We leave this for future work\footnote{We have focused attention in this paper to smooth Ricci flat backgrounds, but since superstring/$M$ theory is consistent in the presence of special kinds of singularities like orbifolds and more general conical singularities \cite{Dixon, bsa2, ew2} it would be interesting to extend the results of this paper to include appropriate singularities.}. Also note that, if we did not quotient by $\alpha$, one would be tempted to say that this metric described the decay of a Kummer $K3$. This is not the case because $T^4/{\bf Z_2}$ is a K3 surface only for the standard spin structure on $T^4$ and in writing the above metric we are explicitly picking an odd spin structure in the covering $T^4$. For general, smooth, Ricci flat $M \cong X/\Gamma^\prime$ without parallel spinors we suspect the existence of an instability of this kind and it would certainly be interesting to pursue this further.

In any case, the conclusions of this section are tentatively very similar to those in \cite{Horowitz,Ooguri} where it was shown that non-supersymmetric quotients in the AdS case tend to be susceptible to generalised Witten instabilities of the kind discussed here.

\section{\it Discussion.}

The questions considered here are related to questions in quantum gravity, such as: {\it is Minkowski spacetime supersymmetric?} \cite{Banks}. 
What are the implications of the existence of {\it stable, compact, simply connected, non-supersymmetric Ricci flat manifolds?}. Well, these would provide stable, non-supersymmetric solutions of superstring/$M$ theory in Minkowski spacetime. A natural presumption is that if such stable, non-supersymmetric, compact Ricci flat manifolds existed, then there would likely be many more of them than the special holonomy cases. In this case, the string landscape would be predominantly non-supersymmetric. From a generic point of view then, one would expect supersymmetry to play virtually no role atall in low energy physics such as the Higgs sector of the Standard Model or dark matter. 
Now suppose that, somewhere in the future, supersymmetric particles such as gluinos, squarks or selectrons are discovered in a collider experiment or elsewhere. Then, if our conjecture is false, we could be left having to explain why nature picks a Ricci flat special holonomy space to model the extra dimensions versus all the other presumably {\it vast} number of possibilities. On the other hand, if the only possible Minkowski spacetimes are actually supersymmetric, this would be a much more satisfying, generic conclusion. This argument makes sense irrespective of the naturalness or hierarchy problem and is consistent if the masses of the newly discovered supersymmetric particles are below the compactification scale.

\bigskip
\large
\noindent
{\bf {\sf Acknowledgements.}}
\normalsize

I would like to thank D. Baraglia, D. Crowley, S. Donaldson, M. Haskins, J. Lotay, C. LeBrun, K. Narain, A. Sen, E. Svanes for comments and suggestions during the course of this work. I especially thank C. Vafa for discussions and encouraging me to write up these results some time ago. The work of BSA is supported by a grant from the Simons Foundation (\#488569, Bobby Acharya)


\bigskip

\begin{thebibliography}{10}


\bibitem{bsa2} B.S. Acharya, {\it On Realising N=1 super Yang-Mills in M theory.} [arXiv:hep-th/0011089]
\bibitem{ew2} B.S. Acharya and E. Witten,
{\it Chiral Fermions from Manifolds of $G_2$-holonomy,} [arXiv:hep-th/0109152].

\bibitem{Banks} T. Banks, {\it Cosmological breaking of supersymmetry?,}
  Int.\ J.\ Mod.\ Phys.\ A {\bf 16} (2001) 910
  doi:10.1142/S0217751X01003998
  [hep-th/0007146].

T.~Banks,
  {\it TASI Lectures on Holographic Space-Time, SUSY and Gravitational Effective Field Theory,}
  arXiv:1007.4001.

\bibitem{Baraglia} D. Baraglia, {\it The alpha invariant of complete intersections,} arXiv:2002.06750

\bibitem{Becker}
  K.~Becker and M.~Becker,
  {\it M theory on eight manifolds,}
  Nucl.\ Phys.\ B {\bf 477} (1996) 155

\bibitem{Berger} Berger, Marcel {\it Sur les groupes d'holonomie homogÃ¨ne des variÃ©tÃ©s Ã  connexion affine et des variÃ©tÃ©s riemanniennes.} (French) Bull. Soc. Math. France 83 (1955), 279â330.
\bibitem{BK} Brendle, S. and Kapouleas, N. (2017), {\it Gluing Eguchi-Hanson Metrics and a Question of Page.} Comm. Pure Appl. Math., 70: 1366-1401. doi:10.1002/cpa.21678arXiv:1405.0056

%\cite{Candelas:1985en}

\bibitem{Bochner} S. Bochner, {\it Vector fields and Ricci curvature}, Bull. Amer. Math. Soc. 52 (1946) 776-797.

\bibitem{CHSW}
  P.~Candelas, G.~T.~Horowitz, A.~Strominger and E.~Witten,
  {\it Vacuum Configurations for Superstrings,}
  Nucl.\ Phys.\ B {\bf 258} (1985) 46.
  doi:10.1016/0550-3213(85)90602-9
  %%CITATION = doi:10.1016/0550-3213(85)90602-9;%%
  %2665 citations counted in INSPIRE as of 31 Aug 2018

\bibitem{Cheeger-Gromoll} Cheeger, Jeff; Gromoll, Detlef. {\it The splitting theorem for manifolds of nonnegative Ricci curvature.} J. Differential Geom. 6 (1971), no. 1, 119--128. doi:10.4310/jdg/1214430220. https://projecteuclid.org/euclid.jdg/1214430220

\bibitem{Dai} Xianzhe Dai, Xiaodong Wang, Guofang Wei, {\it 
On the Stability of Riemannian Manifold with Parallel Spinors,} Inv. Math. 161, 151-176 (2005)

\bibitem{Denef}
  F.~Denef and M.~R.~Douglas,
  {\it Distributions of flux vacua,}
  JHEP {\bf 0405} (2004) 072

\bibitem{Douglas}
  M.~R.~Douglas,
  {\it The Statistics of string / M theory vacua,}
  JHEP {\bf 0305} (2003) 046
M.~R.~Douglas,
  {\it Statistical analysis of the supersymmetry breaking scale,}
  hep-th/0405279.

\bibitem{Dixon}
  L.~J.~Dixon, J.~A.~Harvey, C.~Vafa and E.~Witten,
  {\it ``Strings on Orbifolds,}
  Nucl.\ Phys.\ B {\bf 261} (1985) 678.

\bibitem{Donaldson} S. Donaldson, {\it An Application of Gauge Theory to Four Dimensional Topology}, Journal of Differential Geometry, 18 (1983) (2): 279-315

\bibitem{DonaldsonGluing} S. Donaldson, {\it Calabi-Yau metrics on Kummer surfaces as a model glueing problem,} https://arxiv.org/abs/1007.4218

\bibitem{Fischer-Wolf} Fischer, Arthur E.; Wolf, Joseph A. The structure of compact Ricci-flat Riemannian manifolds. J. Differential Geom. 10 (1975), no. 2, 277--288.

\bibitem{NonSusyStrings}
R.~Rohm,
  {\it Spontaneous Supersymmetry Breaking in Supersymmetric String Theories,}
  Nucl.\ Phys.\ B {\bf 237} (1984) 553.
C.~Kounnas and M.~Porrati,
  {\it Spontaneous Supersymmetry Breaking in String Theory,}
  Nucl.\ Phys.\ B {\bf 310} (1988) 355.
S.~Kachru, J.~Kumar and E.~Silverstein,
  {\it Vacuum energy cancellation in a nonsupersymmetric string,}
  Phys.\ Rev.\ D {\bf 59} (1999) 106004
  A.~Font and A.~Hernandez,
  {\it Nonsupersymmetric orbifolds,}
  Nucl.\ Phys.\ B {\bf 634} (2002) 51
 % doi:10.1016/S0550-3213(02)00336-X
  [hep-th/0202057].

\bibitem{Freedman} M.H. Freedman, {\it The topology of four dimensional manifolds,}  Journal of Differential Geometry, 17 (1982) (3) 357-453

\bibitem{Futaki} A. Futaki, {\it Scalar-flat closed manifolds not admitting positive scalar curvature metrics} , Inv. Math. 112 (1993) 22-30.

\bibitem{GKP} 
  S.~B.~Giddings, S.~Kachru and J.~Polchinski,
  {\it Hierarchies from fluxes in string compactifications,}
  Phys.\ Rev.\ D {\bf 66}, 106006 (2002)


\bibitem{Hitchin}
N. Hitchin, {\it Harmonic Spinors}, Adv. Math. {\bf 14} 1974, 1-55.

\bibitem{HitchinThorpe}
J. Thorpe, {\it Some remarks on the Gauss-Bonnet formula}, J. Math. Mech. 18 (1969) pp. 779--786.
N. Hitchin, {\it On compact four-dimensional Einstein manifolds}, J. Diff. Geom. 9 (1974) pp. 435--442.

\bibitem{Horowitz}
  G.~T.~Horowitz, J.~Orgera and J.~Polchinski,
  {\it Nonperturbative Instability of AdS(5) x S**5/Z(k),}
  Phys.\ Rev.\ D {\bf 77} (2008) 024004
  [arXiv:0709.4262 [hep-th]].
%\cite{Hull:1986kz}

\bibitem{Hull-Strominger}
  C.~M.~Hull,
 {\it Compactifications of the Heterotic Superstring,}
  Phys.\ Lett.\ B {\bf 178} (1986) 357.
  %%CITATION = doi:10.1016/0370-2693(86)91393-6;%%
  %176 citations counted in INSPIRE as of 17 Oct 2018
A.~Strominger,
 {\it Superstrings with Torsion,}
  Nucl.\ Phys.\ B {\bf 274} (1986) 253.
  %%CITATION = doi:10.1016/0550-3213(86)90286-5;%%
  %603 citations counted in INSPIRE as of 17 Oct 2018

\bibitem{dominic} D.D. Joyce, {\it Compact Manifolds of Special Holonomy,} Oxford University Press 2000.

\bibitem{Kapovitch-Lott} N. Kapovitch and J. Lott, {\it On non-collapsed, almost Ricci flat four-manifolds}, arXiv:1606.09041, to appear in American Journal of Mathematics


\bibitem{GregNeil}
  N.~D.~Lambert and G.~W.~Moore,
  {\it Distinguishing off-shell supergravities with on-shell physics,}
  Phys.\ Rev.\ D {\bf 72} (2005) 085018


\bibitem{LeBrun} C. LeBrun, â{\it Four manifolds without Einstein metrics}â
Math. Res. Lett. 3 (1996) 133-147

\bibitem{LeBrun-Singer} C. LeBrun and M. Singer, {\it A Kummer-type construction of self-dual 4-manifolds}, Math. Ann. 300, 165-180 (1994)

\bibitem{Lichnerowicz} A. Lichnerowicz, {\it Spineurs harmonique}, C.R. Acad. Sci. Paris Ser.A-B 257 (1963), 7-9.

\bibitem{alpha} J. Milnor, {\it Remarks concerning spin manifolds, Differential and Combinatorial Topol- ogy }(A symposium in honor of Marston Morse), Princeton Univ. Press, (1965), 55-62.

%\cite{Ooguri:2017njy}
\bibitem{Ooguri}
  H.~Ooguri and L.~Spodyneiko,
  {\it New Kaluza-Klein instantons and the decay of AdS vacua,}
  Phys.\ Rev.\ D {\bf 96} (2017) no.2,  026016
  doi:10.1103/PhysRevD.96.026016
  [arXiv:1703.03105 [hep-th]].
  %%CITATION = doi:10.1103/PhysRevD.96.026016;%%
  %7 citations counted in INSPIRE as of 17 Oct 2018

\bibitem{AdSSwamp} H. Ooguri and C. Vafa, {\it Non-supersymmetric AdS and the Swampland,}
  Adv.\ Theor.\ Math.\ Phys.\  {\bf 21} (2017) 1787


\bibitem{Page} D.N. Page, {\it A periodic but nonstationary gravitational instanton}, Physics Letters B100, 313-315 (1981)

\bibitem{Pfaeffle}
F. Pf{a}ffle, {\it The Dirac spectrum of Bieberbach manifolds}, J. Geom. Phys. 35
(367-385), 2000

\bibitem{SchoenYau}Schoen, Richard; Yau, Shing Tung. {\it On the proof of the positive mass conjecture in general relativity.}  Comm. Math. Phys. 65 (1979), no. 1, 45--76.
Schoen, Richard; Yau, Shing Tung. {\it Proof of the positive mass theorem. II.} Comm. Math. Phys. 79 (1981), no. 2, 231--260. 

\bibitem{Simons} J. Simons, {\it On the Transitivity of Holonomy Systems,}
Annals of Mathematics
Second Series, Vol. 76, No. 2 (Sep., 1962), pp. 213-234

\bibitem{Topiwala} Topiwala, P. {\it A new proof of the existence of K{\"a}hler-Einstein metrics on K3 I,II} Inventiones Math. 89 425-454 (1987)

\bibitem{Wang} Wang, M.Y., {\it Parallel Spinors and Parallel Forms,} Ann Glob Anal Geom (1989) 7: 59. https://doi.org/10.1007/BF00137402

\bibitem{Swampland} C. Vafa, {\it The String landscape and the swampland,}
  hep-th/0509212., 


\bibitem{WittenStability}
E. Witten, {\it A New Proof of the Positive Mass Theorem,} Comm. Math. Phys. 80 (1981) 381-402
\bibitem{SW}
  E.~Witten,
  {\it ``Monopoles and four manifolds,}
  Math.\ Res.\ Lett.\  {\bf 1} (1994) 769
  doi:10.4310/MRL.1994.v1.n6.a13
  [hep-th/9411102].

\bibitem{BubbleofNothing}
  E.~Witten,
  {\it Instability of the Kaluza-Klein Vacuum,}
  Nucl.\ Phys.\ B {\bf 195} (1982) 481.
  %doi:10.1016/0550-3213(82)90007-4
  %%CITATION = doi:10.1016/0550-3213(82)90007-4;%%

\bibitem{ew3} E. Witten, {\it Deconstruction, $G_2$-holonomy and Doublet-Triplet Splitting,}
[arXiv:hep-ph/0201018].
%\bibitem{jeff} J. Harvey and G. Moore, 
%{\sf Superpotentials and Membrane Instantons,}
%[arXiv:hep-th/9907026].
. 


\bigskip

\end{thebibliography}
\end{document}